\documentclass[superscriptaddress,twocolumn,english,floatfix,aps,prd,amsmath,amssymb,longbibliography,10pt]{revtex4-1}
\usepackage[T1]{fontenc}
\usepackage[latin9]{inputenc}
\usepackage{color}
\usepackage{times}
\usepackage{babel}
\usepackage{amsmath}
\usepackage{amssymb}
\usepackage{graphicx}
\usepackage[colorlinks=true]{hyperref}
\hypersetup{citecolor=blue,linkcolor=blue,urlcolor=blue}

\begin{document}

\title{Motion induced by asymmetric excitation of the quantum vacuum}

\author{Jeferson Danilo L. Silva}
\email{jdanilo@ufpa.br}
\affiliation{Campus Salin\'{o}polis, Universidade Federal do Par\'{a}, 68721-000, Salin\'{o}polis,
Brazil}

\author{Alessandra N. Braga}
\email{alessandrabg@ufpa.br}
\affiliation{Instituto de Estudos Costeiros, Universidade Federal do Par\'{a}, 68600-000,
Bragan\c{c}a, Brazil}

\author{Andreson L. C. Rego}
\email{andresonlcr@gmail.com}
\affiliation{Instituto de Aplica\c{c}\~{a}o Fernando Rodrigues da Silveira, Universidade
do Estado do Rio de Janeiro, 20261-232, Rio de Janeiro, Brazil}

\author{Danilo T. Alves}
\email{danilo@ufpa.br}
\affiliation{Faculdade de F\'{i}sica, Universidade Federal do Par\'{a}, 66075-110, Bel\'{e}m,
Brazil}

\date{\today}

\begin{abstract}
During the last fifty-one years, the effect of excitation of the quantum vacuum 
field induced by its coupling with a moving object has been systematically studied.
Here, we propose and investigate a somewhat inverted setting:
an object, initially at rest, whose motion becomes induced by an excitation of the quantum vacuum caused by the object itself. 
In the present model, this excitation occurs asymmetrically on different sides of the object by a variation in time of one of its characteristic parameters, which couple it with the quantum vacuum field.
\end{abstract}
\maketitle
\section{Introduction}
In 1969, Gerald T. Moore published in his PhD thesis the prediction that a mirror in movement can excite the
quantum vacuum, generating photons \cite{Moore-1970}. 
This effect is nowadays known as the dynamical Casimir effect (DCE) and was investigated, during the 1970s, in other
pioneering articles by DeWitt \cite{DeWitt-1975}, Fulling and Davies 
\cite{Fulling-Davies-1976,Davies-Fulling-1977}, Candelas and Deutsch \cite{Candelas-1977}, 
among others.
Since then, many other authors have dedicated to investigate the DCE 
(some excellent reviews on the DCE can be found in Ref. \cite{Dodonov-2009,Dodonov-2010,Dalvit-et-al-2011-CasimirPhysics,Dodonov-2020}).

In his pioneering work, Moore remarked that ``to practical experimental situations, the creation
of photons from the zero-point energy is altogether negligible'' \cite{Moore-1970}.
In an attempt to overcome this difficulty, several ingenious proposals have been made aiming to observe the particle creation from vacuum by experiments based on the mechanical motion of a mirror \cite{Kim-PRL-2006,Brownell-JPA-2008,Motazedifard-2018,Sanz-Quantum-2018,Qin-PRA-2019,Butera-PRA-2019} , 
but this remains as a challenge \cite{Dodonov-2020}.
However, the particle creation from the vacuum occurs, in general, when a quantized field is submitted to time-dependent boundary conditions, with moving mirrors being a particular case. 
Therefore, it is not necessary to move a mirror to generate real particles from the vacuum. 
Within this more general view, alternative ways to detect particle creation from vacuum 
were inspired in the ideas of Yablonovitch \cite{Yablonovitch-PRL-1989} and Lozovik \textit{et al.} \cite{Lozovik-PZhETF-1995}, 
which consist in exciting the vacuum field by means of time-dependent boundary conditions imposed to the field
by a motionless mirror whose internal properties rapidly vary in time. 
In this context, Wilson \textit{et al.} \cite{Wilson-Nature-2011} observed experimentally the particle creation from vacuum,
using a time-dependent magnetic flux applied in a coplanar waveguide (transmission line) with a superconducting quantum interference device
(SQUID) at one of the extremities, changing the inductance of the SQUID, and thus yielding a time-dependent boundary condition \
\cite{Wilson-Nature-2011}. Other experiments have also been done \cite{Lahteenmaki-2013, Vezzoli-2019,Schneider-et-al-PRL-2020},
and other have been proposed \cite{Braggio-2005, Braggio-2008, Braggio-2009,Johansson-PRL-2009, Dezael-Lambrecht-2010, Kawakubo-Yamamoto-2011,Faccio-Carusotto-2011,Naylor-2012,Naylor-2015,Motazedifard-2015}.

When one moves an object imposing changes in the vacuum field, the latter offers resistance, extracting 
kinetic energy from the object, which is converted into real particles. 
In this case, one can say that the net action of the vacuum is against the motion.
Here, we propose and investigate a somewhat inverted situation: 
an object, initially at rest, isolated from everything and just interacting with the quantum vacuum, whose motion becomes induced by an excitation of the vacuum field caused by the object itself.
In other words, we are looking for a situation where the vacuum field acts in favor of a motion in a preferred direction.
With this in mind, we propose a model where an object imposes a change to the quantum field by the time variation
of the properties of the object. 
Resisting to this change, the vacuum field extracts energy from the object,
exciting the quantum field and converting the energy into real particles. 
A fundamental point of the model presented here is that, to get in motion in a preferred direction, the object has to excite 
the quantum vacuum differently on each side, which requires that an asymmetry must be introduced in the object.
Taking into account the same simplified one-dimensional model considered in the pioneering articles on the DCE 
\cite{Moore-1970, DeWitt-1975,Fulling-Davies-1976,Davies-Fulling-1977,Candelas-1977},
we consider the coupling of a static point object with a quantum real scalar field in (1+1)D
via a $\delta-\delta^{\prime}$ potential, which simulates a partially reflecting object with asymmetric
scattering properties on each side \cite{Castaneda-Guilarte-2015,Silva-Braga-Alves-2016}. 
When the coupling parameters vary in time, this model simulates an object exciting
asymmetrically the fluctuations of the quantum vacuum, which produces a non-null mean force acting on the object, 
so that it can get in motion.
%
Then, instead of against, the vacuum acts in favor of the motion in a preferred direction.
But, not completely in favor, since, once in motion, a dynamical Casimir force acts on the object,
so that part of its kinetic energy is extracted by the vacuum fluctuations and goes to the field.

\section{The initial model}

We are interested in an object whose interaction with the field is described by
an asymmetric scattering matrix, intending to excite asymmetrically the quantum vacuum fluctuations. 
The interaction between the object and the field described by a Dirac $\delta$ potential
produces a (left-right) symmetric scattering matrix \cite{Barton-Calogeracos-1995,Calogeracos-Barton-1995}. 
Therefore, to generate an asymmetry, we also consider the presence of an odd $\delta^{\prime}$
term in the description of the interaction, so that our starting point is the following lagrangian 
for a real massless scalar field in (1+1)D:
\begin{equation}
\mathcal{L} = \mathcal{L}_0-[\mu(t)\delta(x)+\lambda_{0}\delta^{\prime}(x)]\phi^{2}(t,x),
\label{model}
\end{equation}
where $\mathcal{L}_0=[\partial_{t}\phi(t,x)]^{2}-[\partial_{x}\phi(t,x)]^{2}$ 
is the lagrangian for the free field, and the real parameters $\lambda_{0}$ and $\mu$, together with $\delta$
and $\delta^{\prime}$ functions, describe the coupling between the
quantum field and a static object located at the point $x=0$. 
These coupling parameters are related to the properties of the object. The
parameter $\mu$ is a prescribed function of time: $\mu(t)=\mu_{0}[1+\epsilon f(t)]$,
where $\mu_{0}\geq0$ is a constant, $f(t)$ is an arbitrary function
such that $|f(t)|\le1$ and $\epsilon\ll1$. In this way, the parameter
$\mu$ is a perturbation in time around the value $\mu_{0}$. 
We consider that this time variation of $\mu$ occurs at the expense
of the internal energy of the object, with null heat transfer between
the object and the environment. 
As we discuss next, the point object described by this model is partially reflective, 
so that a modification in the parameter $\mu$ implies a change in the transmission 
and reflection coefficients. 
Hereafter, we consider that $c=\hbar=1$, tilde indicates
the Fourier transform, and the subscript $+\;(-)$ indicates the right (left) side.

\begin{figure}[h!]
\begin{center}
\includegraphics[width=0.99\columnwidth]{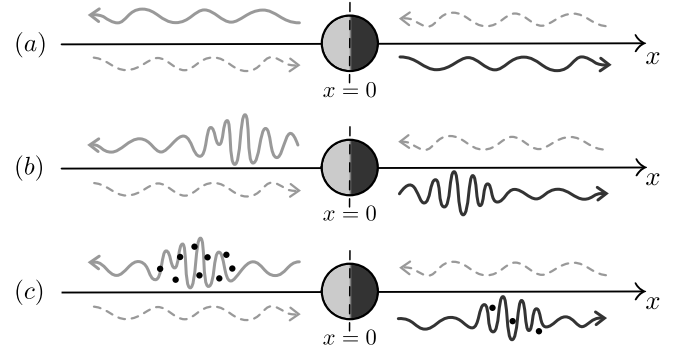}
\end{center}
\caption{Illustration of the excitation of the quantum vacuum
by an object fixed at $x=0$, whose asymmetry in its scattering matrix is represented by its two faces
in gray and black.
The dashed wavy lines represent the unperturbed ``in'' fields
$\varphi_{\text{in}}(t-x)$ (left) and $\psi_{\text{in}}(t+x)$ (right).
(a) Object for $t<-\tau$ ($\tau>0$), when $\mu(t)\approx\mu_0$. 
The solid-gray and solid-dark wavy lines represent the unperturbed ``out'' fields 
$\psi_{\text{out}}(t+x)$ (left) and $\varphi_{\text{out}}(t-x)$ (right), respectively.
(b) Object at an instant $t$, with its parameter $\mu$ varying in time. 
The irregular parts of the solid wavy lines represent the perturbed  parts of the ``out'' fields 
$\psi_{\text{out}}(t+x)$ and $\varphi_{\text{out}}(t-x)$, respectively.
(c) The object for $t>\tau$, when $\mu(t)\approx\mu_0$, and we calculate the number of created particles (represented
by the dark points).
Note that we have more particles produced in the left side.
}
\label{fig-estatica}
\end{figure}

The field equation for this model is 
$(\partial_{t}^{2}-\partial_{x}^{2})\phi(t,x)+2\left[\mu(t)\delta(x)+\lambda_{0}\delta^{\prime}(x)\right]\phi(t,x)=0$.
It will be convenient to split the field as 
$\phi(t,x)=\Theta(x)\phi_{+}(t,x)+\Theta(-x)\phi_{-}(t,x),$
where $\Theta(x)$ is the Heaviside step function and the fields $\phi_{+}$ and $\phi_{-}$ are the sum of two freely counterpropagating fields, namely 
$\phi_{+(-)}(t,x)=\varphi_{\text{out}(\text{in})}(t-x)+\psi_{\text{in}(\text{out})}(t+x)$,
where the labels ``in'' and ``out'' indicate, respectively, the incoming and
outgoing fields with respect to the object (see Fig. \ref{fig-estatica}.a). 
In terms of the Fourier transforms, we can write 
$\phi_{+(-)}(t,x)=\int\frac{\mathrm{d}\omega}{2\pi}\tilde{\phi}_{+(-)}(\omega,x)\text{e}^{-i\omega t}$,
where 
$\tilde{\phi}_{+(-)}(\omega,x) = \tilde{\varphi}_{\text{out}(\text{in})}(\omega)\text{e}^{i\omega x} + 
\tilde{\psi}_{\text{in}(\text{out})}(\omega)\text{e}^{-i\omega x}$.
From the field equation, we get the matching conditions
$\tilde{\phi}(\omega,0^{+})=[(1+\lambda_{0})/(1-\lambda_{0})]\tilde{\phi}(\omega,0^{-})$ 
and 
$\partial_{x}\tilde{\phi}(\omega,0^{+})=[(1-\lambda_{0})/(1+\lambda_{0})]\partial_{x}\tilde{\phi}(\omega,0^{-})+[2/(1-\lambda_{0}^{2})]\int\frac{\mathrm{d}\omega^{\prime}}{2\pi}\tilde{\mu}(\omega-\omega^{\prime})\tilde{\phi}(\omega^{\prime},0^{-})$. 
After an algebraic manipulation in these equations, 
we obtain 
\begin{eqnarray}
\Phi_{\text{out}}(\omega) & = & S_{0}(\omega)\Phi_{\text{in}}(\omega)+\intop\frac{\mathrm{d}\omega^{\prime}}{2\pi}\delta S_{1}(\omega,\omega^{\prime})\Phi_{\text{in}}(\omega^{\prime})\nonumber \\
 &  & +\intop\frac{\mathrm{d}\omega^{\prime}}{2\pi}\int\frac{\mathrm{d}\omega^{\prime\prime}}{2\pi}\delta S_{2}(\omega,\omega^{\prime},\omega^{\prime\prime})\Phi_{\text{in}}(\omega^{\prime\prime}),
\label{eq:out-in}
\end{eqnarray}
with
\begin{equation}
\Phi_{\text{out}(\text{in})}(\omega)=\left(\begin{array}{c}
\tilde{\varphi}_{\text{out}(\text{in})}(\omega)\\
\tilde{\psi}_{\text{out}(\text{in})}(\omega)
\end{array}\right),\, S_{0}(\omega)=\left(\begin{array}{cc}
s_{+}(\omega) & r_{+}(\omega)\\
r_{-}(\omega) & s_{-}(\omega)
\end{array}\right),\nonumber
\end{equation}
where $S_{0}(\omega)$ is the scattering matrix, with 
$s_{\pm}(\omega)=\omega(1-\lambda_{0}^{2})/[i\mu_{0}+\omega(1+\lambda_{0}^{2})]$ and $r_{\pm}(\omega)
=-(i\mu_{0}\mp2\omega\lambda_{0})/[i\mu_{0}+\omega(1+\lambda_{0}^{2})]$
being the transmission and reflection coefficients, respectively. Notice
that the change $\lambda_{0}\leftrightarrow-\lambda_{0}$ leads to $r_{+}(\omega)\leftrightarrow r_{-}(\omega)$,
i.e. the object shifts its properties from one side to the other. 
Moreover, $S_0(\omega)$ is analytic for $\mathrm{Im}\,\omega>0$ (as required by causality \cite{Jaekel-Reynaud-Quant-Opt-1992,Lambrecht-Jaekel-Reynaud-PRL-1996}), unitary and real in the temporal domain. 
The terms $\delta S_{1}(\omega,\omega^{\prime \prime})$ and $\delta S_{2}(\omega,\omega^{\prime},\omega^{\prime\prime})$  represent the first-order and second-order corrections to $S_{0}(\omega)$ due to 
the time-dependence of $\mu$ via $f(t)$.
They are given by 
$\delta S_{1}(\omega,\omega^{\prime})=\epsilon\alpha(\omega,\omega^{\prime})\mathbb{S}(\omega^{\prime})$ 
and 
$\delta S_{2}(\omega,\omega^{\prime},\omega^{\prime\prime})=\epsilon^{2}\alpha(\omega,\omega^{\prime})\alpha(\omega^{\prime},\omega^{\prime\prime})\mathbb{S}(\omega^{\prime\prime})$,
where 
$\alpha(\omega,\omega^{\prime})=-i\mu_{0}\tilde{f}(\omega-\omega^{\prime})/[i\mu_{0}+\omega(1+\lambda_{0}^{2})]$ 
and
\begin{eqnarray}
\mathbb{S}(\omega)=\left(\begin{array}{cc}
s_{+}(\omega) & 1+r_{+}(\omega)\\
1+r_{-}(\omega) & s_{-}(\omega)
\end{array}\right).
\end{eqnarray}
Particularly, $\mu_{0} \rightarrow \infty$ leads to the case of a perfectly reflecting object [$s_{\pm}(\omega)\rightarrow 0$] 
imposing to the field the Dirichlet boundary condition in both sides, for which $\delta S_{1}\rightarrow 0$ and $\delta S_{2}\rightarrow 0$, recovering the configuration of a perfectly reflecting object whose properties do not vary in time.
On the other hand, the limit $\lambda_{0} \rightarrow 1$ 
($\lambda_{0} \rightarrow -1$) also leads to a perfectly reflecting object, but
imposing to the field the Dirichlet and Robin (Robin and Dirichlet) boundary conditions
at the left and right sides of the object, respectively.

\section{Spectrum, energy and momentum}
Let us consider the initial situation ($t<-\tau$) ($\tau>0$) when  
the characteristic parameters of the object are constant $[\lambda_0$ and $\mu(t<-\tau)\approx\mu_0]$ 
and the state of the field is the quantum vacuum (see Fig. \ref{fig-estatica}.a). 
At a certain instant $-\tau$, the properties of the object
start to vary [$\mu_0\rightarrow\mu(t)$], changing the boundary conditions imposed to the field,
exciting the fluctuations of the quantum vacuum in the interval $-\tau <t <\tau$ (see Fig. \ref{fig-estatica}.b).
The final situation ($t>\tau$) is when the object recovers its constant characteristic parameters 
$[\lambda_0$ and $\mu(t>\tau)\approx\mu_0]$ and real particles are created (see Fig. \ref{fig-estatica}.c).
The spectrum of created particles can be computed by 
$n(\omega) = 2\omega\,\mathrm{Tr} 
\langle 0_{\text{in}}|\Phi_{\text{out}}(-\omega)\Phi_{\text{out}}^{\mathrm{T}}(\omega)|0_{\text{in}}\rangle$
\cite{Lambrecht-Jaekel-Reynaud-PRL-1996}. From Eq. (\ref{eq:out-in}), calculated at order up to $\mathcal{O}(\epsilon^{2})$, we have that 
$n(\omega)=n_{+}(\omega) + n_{-}(\omega)$,
where 
\begin{equation}
n_{\pm}(\omega)=\frac{\epsilon^{2}}{2\pi^2}(1\pm\lambda_{0})^{2}(1+\lambda_{0}^{2})\int_{0}^{\infty}{\mathrm{d}\omega^{\prime}}\eta(\omega,\omega^{\prime}),
\label{eq:Npm}
\end{equation}
with $\eta(\omega,\omega^{\prime})=\Upsilon(\omega)\Upsilon(\omega^{\prime})|\tilde{f}(\omega+\omega^{\prime})|^{2}$ and
$\Upsilon(\omega)=\mu_{0}\omega/\left[\mu_{0}^{2}+\omega^{2}(1+\lambda_{0}^{2})^{2}\right]$.
Therefore, we get 
$n_{-}(\omega)=[({1-\lambda_{0}})/({1+\lambda_{0}})]^{2}n_{+}(\omega)$,
which means that the spectrum for one side of the object differs from
the other one by a frequency-independent global factor. 
For $\lambda_{0}>0$ ($\lambda_{0}<0$) $n_{-}(\omega)$ is smaller (greater) than $n_{+}(\omega)$.

The total number of created particles is given by 
$\mathcal{N}=\int_{0}^{\infty}\mathrm{d}\omega \,n(\omega)$,
and the number in each side of the object is 
$\mathcal{N}_{\pm}=\int_{0}^{\infty}\mathrm{d}\omega \,n_{\pm}(\omega)$,
so that we can write
$\mathcal{N}_{-} = 
[({1-\lambda_{0}})/({1+\lambda_{0}})]^{2}\mathcal{N}_{+}$.
Note that $\mathcal{N}$ is greater in the
right side of the object if $\lambda_{0}>0$ and smaller if $\lambda_{0}<0$
(this latter case is illustrated in Fig. \ref{fig-estatica}.c).
Particularly, for the perfectly reflecting cases where $\lambda_0= 1$ 
$(\lambda_0= -1)$, we get $\mathcal{N}_{-}=0$ $(\mathcal{N}_{+}=0)$, 
so that the particles are created only in one side of the object. 

The energy and momentum of the created particles in each side are
given, respectively, by 
$\mathcal{E}_{\pm}=\int_{0}^{\infty}\mathrm{d}\omega\,\omega n_{\pm}(\omega)$
and 
$\mathcal{P}_{\pm}=\pm\mathcal{E}_{\pm}$.
The total energy $\mathcal{E}$ and momentum $\mathcal{P}$ are 
$\mathcal{E}=\mathcal{E}_{+}+\mathcal{E}_{-}$
and 
$\mathcal{P}=\mathcal{P}_{+}+\mathcal{P}_{-}$.
Specifically,
\begin{equation}
{\mathcal{P}}=\frac{2\epsilon^{2}}{\pi^2}\lambda_{0}(1+\lambda_{0}^{2})
\int_{0}^{\infty}{\mathrm{d}\omega}\int_{0}^{\infty}{\mathrm{d}\omega^{\prime}}\omega\eta(\omega,\omega^{\prime}),
\label{P-particles}
\end{equation}
which is negative for $\lambda_0<0$ and positive for $\lambda_0>0$.
Then, a static object, initially fixed at $x=0$, 
with its properties varying in time, can excite asymmetrically 
the fluctuations of the quantum vacuum, generating into the field
a net momentum $\mathcal{P}\neq 0$. 
For instance, for the perfectly reflecting case where $\lambda_0=1$ $(\lambda_0=-1)$, 
we obtain $\mathcal{P}_{-}=0$ $(\mathcal{P}_{+}=0)$, 
so that momentum is transferred to the field (by exciting it) just in one of
the sides of the object. This net momentum implies in a net force acting on the object.

\section{Force on the static object}

Let us now obtain the expression for the mean force acting on the object 
at $x=0$ due to the field fluctuations (see Fig. \ref{fig-estatica}.b).
The components of the energy-momentum tensor for a scalar field in
$1+1$ dimensions are given by  
$T_{00}=T_{11}=\left[\varphi^{\prime}(t-x)\right]^{2}+\left[\psi^{\prime}(t+x)\right]^{2}\equiv E(t,x)$,
and $T_{01}=T_{10}=\left[\varphi^{\prime}(t-x)\right]^{2}-\left[\psi^{\prime}(t+x)\right]^{2}\equiv P(t,x)$,
where $E(t,x)$ and $P(t,x)$ are the energy and momentum densities
respectively, and their mean values can be written as 
$\langle E_j(t,x)\rangle$ $=\mathrm{Tr}[\partial_{t}\partial_{t^{\prime}}
\langle \Phi_j(t,x)\Phi_j^{\mathrm{T}}(t^{\prime},x)\rangle]_{t=t^{\prime}}$ 
and 
$\langle P_j(t,x)\rangle$ $=$ $\text{Tr}[\text{diag}(1,-1)
\partial_{t}\partial_{t^{\prime}}\langle \Phi_j(t,x)
\Phi_j^{\mathrm{T}}(t^{\prime},x)\rangle]_{t=t^{\prime}}$, 
where $j={\text{out},\text{in}}$.
The force acting on the object
due to the field fluctuations can be found as the difference between
the radiation pressure ($T_{11}$), on the left and on the right sides of the object. Therefore,
the mean force is given by 
$
F_\mu(t) = \langle P_{\text{in}}(t,0)-P_{\text{out}}(t,0)\rangle$
and it can be written as
\begin{eqnarray}
F_{\mu}(t)\approx F_{\mu}^{(1)}(t) \epsilon + F_{\mu}^{(2)}(t) \epsilon^{2}.
\label{eq:force}
\end{eqnarray}
Taking its Fourier transform and considering
Eq. (\ref{eq:out-in}), we obtain 
$\tilde{F}_{\mu}(\omega)\approx\tilde{F}_{\mu}^{(1)}(\omega)\epsilon + \tilde{F}_{\mu}^{(2)}(\omega)\epsilon^{2}$, 
where the mean value was taken considering a vacuum as the initial
state of the field. The first-order term is 
$\tilde{F}_{\mu}^{(1)}(\omega)=\chi_{1}(\omega)\tilde{f}(\omega)$ 
where
\begin{equation}
\chi_{1}(\omega)=\frac{2\lambda_{0}\omega^{2}}{\pi\rho^{2}(\lambda_{0}^{2}+1)}\bigg\{\frac{\rho+i}{\rho+2i}[2i\arctan\rho-\ln(\rho^{2}+1)]-i\rho\bigg\},\nonumber
\end{equation}
and 
$\rho=(\lambda_{0}^{2}+1)\omega/\mu_{0}$. 
The second-order term is 
$\tilde{F}_{\mu}^{(2)}(\omega)=\int\mathrm{d}\omega^{\prime}\int\mathrm{d}\omega^{\prime\prime}\chi_{2}(\omega,\omega^{\prime},\omega^{\prime\prime})\alpha(\omega^{\prime},-\omega^{\prime\prime})\alpha(\omega-\omega^{\prime\prime},\omega^{\prime})$
where 
\begin{eqnarray*}
\chi_{2}(\omega,\omega^{\prime},\omega^{\prime\prime}) & = & \lambda_{0}[h(-\omega^{\prime})\varTheta(\omega^{\prime\prime})\omega^{\prime2}(1+\lambda_{0}^{2})-\omega^{\prime\prime}/2]\\
 &  & \times h(\omega^{\prime\prime})(\omega^{\prime\prime}-\omega)\mathrm{sgn}(\omega^{\prime\prime})/\pi^{2},
\end{eqnarray*}
and $h(\omega)=1/[i\mu_{0}+\omega(1+\lambda_{0}^{2})]$.

\section{Free to move}

\begin{figure}[h!]
\begin{center}
\includegraphics[width=0.99\columnwidth]{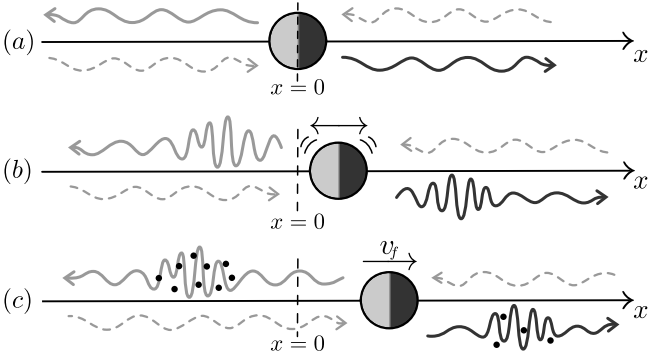}
\end{center}
\caption{Illustration of 
an object, initially at rest (a) but free to move [(b) and (c)], 
whose motion becomes induced by an excitation of the quantum vacuum caused by the object itself.
The object, characterized by an asymmetric scattering matrix, is represented by a circle with two faces in gray and black.
The dashed wavy lines represent the unperturbed ``in'' fields
$\varphi_{\text{in}}(t-x)$ (left) and $\psi_{\text{in}}(t+x)$ (right).
(a) The object for $t<-\tau$, when $\mu(t)\approx\mu_0$.  
The solid-gray and solid-dark wavy lines represent the unperturbed ``out'' fields 
$\psi_{\text{out}}(t+x)$ (left) and $\varphi_{\text{out}}(t-x)$ (right), respectively.
(b) The object at an instant $t$, when the parameter $\mu$ is varying in time. 
The irregular parts of the solid-gray and of the solid-dark wavy lines represent the perturbed 
parts of the ``out'' fields $\psi_{\text{out}}(t+x)$ and $\varphi_{\text{out}}(t-x)$, respectively.
Under this situation, the object is subjected to the force
$
{F_{\mu q}^{(1)}(t,\dot{q}(t))\epsilon}
+
{F_{\mu q}^{(2)}(t,\dot{q}(t))\epsilon^{2}}
+
{F_{q}(\partial_{t}^{3}q(t),\partial_{t}^{4}q(t),...)}
$
and gets in motion.
(c) The object for $t>\tau$, when $\mu(t)\approx\mu_0$, and we calculate the number of created particles (represented
by the dark points), moving with a final constant mean velocity $v_f$.
Note that we have more particles produced in the left side, coinciding
in this illustration with a larger flux of particle momentum.
}
\label{fig-dinamica}
\end{figure}

So far, the object has been assumed to be fixed at $x=0$, as described by the langrangian \eqref{model}. 
Now, let us consider that for $t<-\tau$ the object is kept at $x=0$
(Fig. \ref{fig-dinamica}.a), but for $t>-\tau$ it is free to move. 
%
Even if $\mu(t)=\mu_0$ for $t>-\tau$, a fluctuating force from the quantum vacuum field would
act on the object, so that it would start a Brownian motion 
\cite{Sinha-Sorkin-PRB-1992,Jaekel-Reynaud-1993, Stargen-Kothawala-Sriramkumar-PRD-2016,Wang-Zhu-Unruh-PRD-2017}.
On the other hand, one can use one of the degrees of freedom of the model, namely the initial mass of the object $(M_{0})$, 
to simplify the problem. 
Assuming $M_{0}$ sufficiently large, the mean-squared displacement in the position of the object, during the interval 
$-\tau<t<\tau$, can be neglected and, therefore, the object remains at $x\approx 0$ in the mentioned interval.

Now, let us consider again $\mu(t)=\mu_{0}[1+\epsilon f(t)]$ in the interval $-\tau<t<\tau$ 
(Fig. \ref{fig-dinamica}.b), with $M_0$ remaining
large enough to the Brownian motion be neglected, and the object free to move for $t>-\tau$ (Fig. \ref{fig-dinamica}.c).
For this case, the boundary condition imposed to the field on the static object at $x=0$ must now
be replaced by a boundary condition considered in the instantaneous position $x=q(t)$ of the moving object, 
observed from the point of view of an inertial frame where the object is instantaneously at rest (called tangential frame), 
and then be mapped into a boundary condition viewed by the laboratory system 
\cite{Jaekel-Reynaud-Quant-Opt-1992,Maia-Neto-JPA-1994,Rego-Mintz-Farina-Alves-2013,Silva-Braga-Rego-Alves-2015,Silva-Braga-Alves-2016}.
This means that the force $F_{\text{\ensuremath{\mu}}}(t)$ [Eq. \eqref{eq:force}] 
needs to be replaced by a modified 
$F_{\text{\ensuremath{\mu q}}}(t,\dot{q}(t))$, 
which now can depend on the velocity of the object 
[for consistency, we consider that $F_{\text{\ensuremath{\mu q}}}(t,0)=F_{\text{\ensuremath{\mu}}}(t)$].
In addition to the force $F_{\text{\ensuremath{\mu q}}}(t,\dot{q}(t))$,
the motion of the object gives rise to an extra disturbance to the vacuum field, from which arises an
additional (dynamical Casimir) force acting on the object.
This force is here represented by
$F_{\text{\ensuremath{q}}}(\partial_{t}^{3}q(t),\partial_{t}^{4}q(t),...)$,
so that it acts on non-uniformly accelerating objects
(see, for instance, Ref. \cite{Fulling-Davies-1976,Ford-Vilenkin-1982}).
For consistency, we assume that, for a static object, $F_{\text{\ensuremath{q}}}(0,0,...)=0$.
As an example, if one considers, for instance $\lambda_{0}\rightarrow-1$, 
the object imposes to the field on the left (right) side the Robin (Dirichlet) boundary condition. 
For this case, 
$F_{\text{\ensuremath{q}}}(\partial_{t}^{3}q(t),\partial_{t}^{4}q(t),...)\approx{1}/({6\pi})\partial_{t}^{3}q(t)-{1}/({6\pi\mu_{0}})\partial_{t}^{4}q(t)$ \cite{Ford-Vilenkin-1982,Jaekel-Reynaud-Quant-Opt-1992,Mintz-Farina-MaiaNeto-Rodrigues-2006-I}.
In summary, two forces act on the object free to move, a force 
$F_{\text{\ensuremath{\mu q}}}(t,\dot{q}(t))$ 
(related to the time-varying properties of the object) and a force 
$F_{\text{\ensuremath{q}}}(\partial_{t}^{3}q(t),\partial_{t}^{4}q(t),...)$
(related to the motion of the object).

From the energy conservation, and assuming that the excitation of the quantum vacuum occurs 
at the expense of the energy of the object [so that its initial mass $M_{0}$ 
becomes time-dependent: $M_{0}\rightarrow M(t)$], we have
$M(t)\approx M_{0}[1-\mathcal{E_{\text{field}}}(t)/M_{0}]
/[1+\dot{q}(t)^{2}/2]$,
where $\mathcal{E_{\text{field}}}(t)$ 
is the energy stored in the field, and the velocities of the object are considered non-relativistic
(remember that $c=1$).
We consider, now, the approximation 
$\mathcal{E_{\text{field}}}(t)/M_{0}\ll 1$,
which means that the energy $\mathcal{E_{\text{field}}}(t)$ 
is negligible if compared to the initial energy of the object. We also consider the non-relativistic assumption
$\dot{q}(t)^{2}\ll1$.
Then we have $M(t)\approx M_{0}$.

\begin{figure}[h!]
\begin{center}
\includegraphics[width=1\columnwidth]{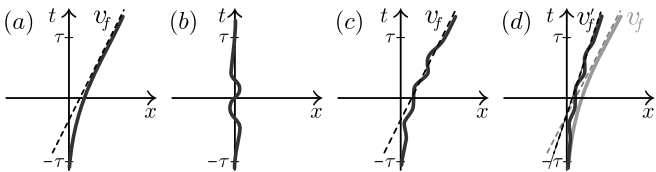}
\end{center}
\caption{Mean trajectories $q(t)$ (continuous lines) of an object initially $(t<-\tau)$ at rest.
(a) Illustration of $q(t)$ when just the term ${F_{\text{\ensuremath{\mu}}}^{(2)}(t)\epsilon^{2}}$
is considered. The mean final velocity $v_f$ is indicated by the dashed line.
(b) Illustration of $q(t)$ when only the term ${F_{\text{\ensuremath{\mu}}}^{(1)}(t)\epsilon}$ is taken into account.
The null contribution of this force to the mean final 
velocity of the object is indicated by the vertical inclination of the curve $q(t)$ for $t>\tau$.
(c) Illustration of $q(t)$ when the sum ${F_{\text{\ensuremath{\mu}}}^{(1)}(t)\epsilon}
+{F_{\text{\ensuremath{\mu}}}^{(2)}(t)\epsilon^2}$ is considered.
The mean trajectory is drawn by introducing deformations on the curve shown in the case (a), but maintaining the same mean final velocity $v_f$.
(d) Situation when all terms are considered:
$
{F_{\text{\ensuremath{\mu}}}^{(1)}(t)\epsilon}
+
{F_{\text{\ensuremath{\mu}}}^{(2)}(t)\epsilon^{2}}
+
F_{\text{\ensuremath{q}}}(\partial_{t}^{3}q(t),\partial_{t}^{4}q(t),...)
$. 
The net dissipation of the kinetic energy by $F_{\text{\ensuremath{q}}}(\partial_{t}^{3}q(t),\partial_{t}^{4}q(t),...)$ 
is indicated by the mean final velocity $v_f^{\prime}$, with $|v_f^{\prime}|<|v_f|$. 
For comparison purposes, the figure also shows, in gray lines, the situation described in (a),
when only $F_{\text{\ensuremath{\mu}}}^{(2)}(t)\epsilon^{2}$ was considered.
}
\label{fig-q-t}
\end{figure}

Considering all forces acting on the object, we have the equation of motion
$F_{\text{\ensuremath{\mu q}}}(t,\dot{q}(t))
+F_{\text{\ensuremath{q}}}(\partial_{t}^{3}q(t),\partial_{t}^{4}q(t),...)
\approx M_{0}\ddot{q}(t)$. 
We also consider that, up to second order in $\epsilon$, the force 
$F_{\text{\ensuremath{\mu q}}}(t,\dot{q}(t))$ 
can be written as
$F_{\text{\ensuremath{\mu q}}}(t,\dot{q}(t))\approx F_{\text{\ensuremath{\mu q}}}^{(1)}(t,\dot{q}(t))\epsilon+F_{\text{\ensuremath{\mu q}}}^{(2)}(t,\dot{q}(t))\epsilon^{2}$,
which is an extension of Eq. \eqref{eq:force}, assuming
$F_{\text{\ensuremath{\mu q}}}^{(1)}(t,0)=F_{\mu}^{(1)}(t)$,
and
$F_{\text{\ensuremath{\mu q}}}^{(2)}(t,0)=F_{\mu}^{(2)}(t)$.
Then, we write the equation of motion as:
$
{F_{\text{\ensuremath{\mu q}}}^{(1)}(t,\dot{q}(t))\epsilon}
+
{F_{\text{\ensuremath{\mu q}}}^{(2)}(t,\dot{q}(t))\epsilon^{2}}
+
F_{\text{\ensuremath{q}}}(\partial_{t}^{3}q(t),\partial_{t}^{4}q(t),...)
\approx
{M_{0}}\ddot{q}(t).
$
Now, one can use two of the degrees of freedom of the model, namely the value of $\epsilon$ 
and the initial mass $M_{0}$, 
to simplify the problem. 
For instance, let us consider the changes
$\epsilon\rightarrow10^{p}\epsilon$,
and
$M_{0}\rightarrow10^{2p}M_{0}$
(with $p>0$) to build a new situation for which the equation 
of motion is 
$
{F_{\mu q}^{(1)}(t,\dot{q}(t))\epsilon}/{10^{p}}
+
{F_{\mu q}^{(2)}(t,\dot{q}(t))\epsilon^{2}}
+
{F_{q}(\partial_{t}^{3}q(t),\partial_{t}^{4}q(t),...)}/{10^{2p}}\approx M_{0}\ddot{q}(t).
$
Increasing the value of $p$, we can inhibit the effect of the first and third terms
in the last equation, so that we can set up a situation where these terms can be neglected in comparison
with the second one, resulting in the approximate equation of motion
$F_{\mu q}^{(2)}(t,\dot{q}(t))\epsilon^{2}\approx M_{0}\ddot{q}(t)$.
In other words, for a suitable choice of the initial mass $M_0$, the force
$F_{\mu q}^{(2)}(t,\dot{q}(t))$ 
defines effectively the mean trajectory of the object.
%
By keeping increasing the mass $M_{0}$, one can produce smaller accelerations, 
so that the velocities of the object are of such magnitude that the boundary condition imposed by the object on the field, 
considered by the tangential frame, 
after mapped into the boundary condition viewed by the laboratory system, can be approximately given by
$F_{\mu q}^{(2)}(t,\dot{q}(t))\approx F_{\mu q}^{(2)}(t,0)=F_{\mu}^{(2)}(t)$.
With this approximation, we have the equation of motion given by
$F_{\mu}^{(2)}(t)\epsilon^{2}\approx M_{0}\ddot{q}(t)$.
Integrating in time, $\int_{-\infty}^{+\infty} F_{\mu}^{(2)}(t) \mathrm{d}t$,
using the mentioned formula shown for $\tilde{F}_{\mu}^{(2)}(\omega)$, 
the property $\eta(\omega,\omega^\prime)=\eta(\omega^\prime,\omega)=\eta(-\omega,-\omega^\prime)$,
and also considering that $\dot{q}(t<-\tau)=0$, we get
\begin{equation}
v_f\approx -{{\cal P}}/{M_{0}},
\label{v-f}
\end{equation}
where $v_f=\dot{q}(t>\tau)$ is the mean final velocity.
Note that $-{\cal P}$ is the opposite of the net
momentum transferred from the object to the field [see Eq. \eqref{P-particles}],
so that the momentum transferred to the object, caused by the action of
the force $F_{\text{\ensuremath{\mu}}}^{(2)}(t)\epsilon^{2}$, is directly correlated with
the particle creation process. This situation is illustrated in Figs. \ref{fig-dinamica}.c
and \ref{fig-q-t}.a.

Let us give attention to another situation. The 
values of $\epsilon$ and $p$ can be chosen in such way that the term related to ${F_{\text{\ensuremath{\mu q}}}^{(1)}(t,\dot{q}(t))}$
becomes dominant in relation to ${F_{\text{\ensuremath{\mu q}}}^{(2)}(t,\dot{q}(t))\epsilon^{2}}$.
We consider [as done in a similar way for ${F_{\text{\ensuremath{\mu q}}}^{(2)}(t,\dot{q}(t))\epsilon^{2}}$] 
the approximation
$F_{\text{\ensuremath{\mu q}}}^{(1)}(t,\dot{q}(t))\approx F_{\text{\ensuremath{\mu q}}}^{(1)}(t,0)=F_{\text{\ensuremath{\mu}}}^{(1)}(t)$.
It can be shown that $\int_{-\infty}^{+\infty} F_{\text{\ensuremath{\mu}}}^{(1)}(t) \mathrm{d}t=0$,
which means that, although this force makes the position of the object
vary in time, the total net momentum transferred to the object is null.
This situation is illustrated in Fig. \ref{fig-q-t}.b. 

Another situation can be obtained by manipulating 
the values of $\epsilon$ and $p$ in such way that
$F_{\text{\ensuremath{\mu}}}^{(1)}(t)\epsilon$ 
and  
$F_{\text{\ensuremath{\mu}}}^{(2)}(t)\epsilon^2$ have similar magnitudes. 
The presence of $F_{\text{\ensuremath{\mu}}}^{(1)}(t)\epsilon$ 
disturbs the mean trajectory but does not change the final velocity $v_f$ obtained if only 
$F_{\text{\ensuremath{\mu}}}^{(2)}(t)\epsilon^2$ was considered.
This case is illustrated in Fig. \ref{fig-q-t}.c.

Finally, we can set up the values of $\epsilon$ and $p$
so that we have to consider  all terms, ${F_{\text{\ensuremath{\mu q}}}^{(1)}(t)}\epsilon$,  ${F_{\text{\ensuremath{\mu q}}}^{(2)}(t)\epsilon^{2}}$ and $F_{\text{\ensuremath{q}}}(\partial_{t}^{3}q(t),\partial_{t}^{4}q(t),...)$.
One can see that $\int_{-\infty}^{+\infty}F_{\text{\ensuremath{q}}}(\partial_{t}^{3}q(t),\partial_{t}^{4}q(t),...)\dot{q}(t)\mathrm{d}t<0$,
so that the net action of $F_{\text{\ensuremath{q}}}(\partial_{t}^{3}q(t),\partial_{t}^{4}q(t),...)$ is to dissipate energy of the object.
This leads to a mean final velocity $v_f^{\prime}$ with a smaller magnitude
if compared with $v_f$ $(|v_f|>|v_f^{\prime}|)$ obtained if only 
$F_{\text{\ensuremath{\mu}}}^{(1)}(t)\epsilon+F_{\text{\ensuremath{\mu}}}^{(2)}(t)\epsilon^2$ was considered.
This situation is illustrated in Fig. \ref{fig-q-t}.d.

\section{Summary of the results and final remarks}
In the model proposed here, a static object, isolated from everything and just interacting with
the quantum vacuum, gets in motion by exciting the vacuum.
Then, the net action of the vacuum field is in favor of the motion, instead of against, as
it occurs in the usual dynamical Casimir effect.
This motion requires a time variation of one of the parameters $[\mu(t)]$ of the object,
which couple it with the quantum vacuum field. 
Resisting to this change, the vacuum field extracts energy from the object,
converting this energy into real particles.
The motion also requires an asymmetrical vacuum excitation on each side,
which can be achieved by an interaction field-object described by an asymmetric scattering matrix.

The mean force acting on the object due to $\mu(t)$ can be divided in two parts, 
the forces $F_{\text{\ensuremath{\mu}}}^{(1)}(t)\epsilon$ and $F_{\text{\ensuremath{\mu}}}^{(2)}(t)\epsilon^2$. 
These forces only exist (are non-null) owing to the asymmetry of the object,
which means that the asymmetry is fundamental to the rise of these quantum forces.
%

The part of the force related to $F_{\text{\ensuremath{\mu}}}^{(1)}(t)$
is a manifestation of the disturbed vacuum field in order $\epsilon$, 
and the correspondent force can remove the object from the rest, but 
gives no contribution to the net momentum.
On the other hand, the term related to $F_{\text{\ensuremath{\mu}}}^{(2)}(t)$
is a manifestation of the disturbed vacuum field in order $\epsilon^2$,
and it is a direct consequence of the momentum transferred to the object by the created particles.

The mean forces, $F_{\text{\ensuremath{\mu}}}^{(1)}(t)\epsilon$ 
and $F_{\text{\ensuremath{\mu}}}^{(2)}(t)\epsilon^2$, described in the present paper,
are quantum forces emerged from the asymmetry and time-varying properties.
In conjunction with the dynamical Casimir force $F_{\text{\ensuremath{q}}}(\partial_{t}^{3}q(t),\partial_{t}^{4}q(t),...)$, 
these three forces define, in the approximation considered here,
a mean trajectory for the object. 
Finally, the object, starting from the rest,
gets a non-null mean final velocity, so that, exciting the vacuum, it moves.


\section{Acknowledgments}

D.T.A. thanks the hospitality of the Centro de F\'{i}sica, Universidade do Minho, Braga, Portugal.

%

\end{document}